\documentclass[aps,prd,amsmath,amssymb,twocolumn,superscriptaddress,preprintnumbers,nofootinbib]{revtex4-1}
\pdfoutput=1
\usepackage{latexsym}
\usepackage{amsthm}
\usepackage{amsmath}
\usepackage{graphicx}
\usepackage{dcolumn}
\usepackage{bm}
\usepackage{textcomp}
\usepackage{amssymb}

\newcommand{\be}{\begin{eqnarray}}
\newcommand{\ee}{\end{eqnarray}}
\newcommand{\bea}{\begin{eqnarray}}
\newcommand{\eea}{\end{eqnarray}}

\newcommand{\GeV}{~\mathrm{GeV}}
\newcommand{\TeV}{~\mathrm{TeV}}

\newcommand{\zp}{Z^{\prime}}
\newcommand{\Ap}{a'}
\newcommand{\hp}{h'}
\newcommand{\Hp}{H'}
\newcommand{\uonep}{U(1)^\prime}
\newcommand{\mzp}{M_{Z^\prime}}
\newcommand{\gzp}{g_{z^\prime}}

\newcommand{\dimEn}{\varepsilon}
\newcommand{\ymax}{y_{\rm max}}
\newcommand{\etahat}{\hat{\eta}}
\newcommand{\boost}{\gamma}
\newcommand{\DeltaRmin}{\Delta R_{\rm min}}

\begin{document}

\title{Multiphotons and Photon-Jets}
\author{Natalia Toro}
\affiliation{Perimeter Institute for Theoretical Physics 31 Caroline St. N, Waterloo, Ontario, Canada N2L 2Y5.}
\author{Itay Yavin}
\affiliation{Perimeter Institute for Theoretical Physics 31 Caroline St. N, Waterloo, Ontario, Canada N2L 2Y5.}
\affiliation{Department of Physics \& Astronomy, McMaster University
1280 Main St. W, Hamilton, Ontario, Canada, L8S 4L8}

\begin{abstract}
We discuss an extension of the Standard Model with a new vector-boson decaying predominantly into a multi-photon final state through intermediate light degrees of freedom. The model has a distinctive phase in which the photons are collimated.  As such, they would fail the isolation requirements of standard multi-photon searches, but group naturally into a novel object, the \emph{photon-jet}. Once defined, the photon-jet object facilitates more inclusive searches for similar phenomena.  
We present a concrete model, discuss photon-jets more generally, and outline some  strategies that may prove useful when searching for such objects.  
\end{abstract}

\maketitle


New heavy vector-bosons have long been discussed in the literature and searched for in colliders~\cite{Langacker:2008yv}. The production of the vector-boson in high-energy collisions is usually searched for by looking for a heavy resonance in a di-fermion final state. In this \textsl{Letter} we discuss the possibility of a vector-boson decaying predominantly into a multi-photon final state and the associated phenomenology.  A case of particular interest is when the vector-boson decays into highly boosted light resonances, which in turn decay to collimated di-photons.  Such di-photon pairs fail standard isolation criteria,  instead giving rise to a novel object that is likely recognizable at colliders, dubbed a ``photon-jet''.   

Theories with new massive vector-bosons usually involve a broken gauge symmetry. 
If the symmetry is broken by charged scalars that higgs the vector-boson, then the vector boson necessarily couples to the un-eaten components of these scalars and can decay into them if kinematically allowed.  This decay dominates if no other light fields are charged under the broken gauge symmetry.  The observation of the heavy vector-boson then depends on the decay of the scalars, as is common in hidden valley models~\cite{Strassler:2006im}.  If these scalars are light and do not have any renormalizable interaction with the SM, then they may decay only into photons through terms of the form $\phi F_{\mu\nu}F^{\mu\nu}$ where $F_{\mu\nu}$ is the electromagnetic field strength.  
Through relativistic kinematics, the mass spectrum of the theory dictates the degree of collimation of these photons. If the scalars are comparable in mass to the vector-boson then the photons will be fairly well isolated and result in multi-photon signature. On the other hand, if the scalars are much lighter than the vector-boson, they are ultra-relativistic so the photons resulting from their decay are tightly collimated (as is commonly seen in $\pi^0$ decay). Such ``photon-jets'' would fail the isolation criteria present in general photon searches. In extreme cases, when the scalars are very light, each of the scalars may be reconstructed as  a single photon. A collimated photon bunch of this sort was previously considered in Ref.~\cite{Dobrescu:2000jt} where the authors examined the possibility of the Higgs boson decay into two light pseudoscalars. 
Amusingly, a massive vector-boson decaying into such narrow photon-jets would be reconstructed as a diphoton decay of a spin-1 particle, in apparent violation the Landau-Yang theorem which forbids such transitions~\cite{Yang:1950rg}. 

A concrete model realizing the above phenomenology includes a new gauge symmetry $\uonep$ and its carrier, $\zp$, two complex scalars, $S_{1,2}$, charged under this symmetry, and a set of heavy fermions $\psi,\psi^c$ and $\chi, \chi^c$ that are charged both under the new group as well as hypercharge as $(\pm2,\pm1)$ and $(\mp1,\mp1)$, respectively. The high energy Lagrangian is,
\be
\mathcal{L} &=& \mathcal{L}_{\uonep} + \mathcal{L}_{\rm fermions} -\frac{\epsilon}{2}\zp_{\mu\nu}B^{\mu\nu} \\ \nonumber
\mathcal{L}_{\uonep} &=&  -\tfrac{1}{4}\zp_{\mu\nu}Z^{'\mu\nu} +\sum_{i=1,2} \left| D_\mu S_i \right|^2 + V(S_1,S_2)\\ \nonumber
\mathcal{L}_{\rm fermions} &=& \text{kinetic terms} + y_1  S_1 \psi \chi + y_2 S_2 \psi^c\chi^c
\ee
The fermions will generically drive the scalars to develop a vacuum expectation value and break the $\uonep$. This will render the vector-boson massive, with mass $\mzp^2 = \gzp^2\left(\langle S_1\rangle^2 + \langle S_2\rangle^2 \right)$ where $\langle S_i\rangle$ is the expectation value of $\sqrt{2} S_i$. The fermions are also rendered massive with masses $y_1\langle S_1\rangle$ and $y_2\langle S_2\rangle$ for $\psi\chi$ and $\psi^c\chi^c$, respectively.
The separate  global phase symmetries on $S_1$ and $S_2$ result in two Goldstone bosons at low energies~\cite{Axion}. 
One combination becomes the longitudinal component of the $\uonep$ through the Higgs mechanism. The remaining physical pseudoscalar, which we denote by $\Ap$,  may be naturally light with a mass sensitive only to terms that break the separate phase symmetry of the two complex scalars. The masses of the rest of the scalars are determined by the detailed quartic couplings of the scalar potential $V(S_1,S_2)$. We note that the matter content above is anomaly free, but the sum of $\uonep\times U_Y(1)$ charges does not vanish, giving rise to a logarithmic running of kinetic mixing. This can be cancelled by the introduction of a heavy Dirac fermion of charge $(-3,1)$ with an arbitrary mass $M$. The kinetic mixing term in this model is then given by $\epsilon = \epsilon_{UV} + (3 \gzp g_Y/16\pi^2)\sum_i \log(y_i\langle S_i\rangle/M)$, which we treat as a free parameter.  This mixing results in a $\zp$ which interpolates between coupling dominantly to the hypercharge current when $\mzp \gg M_{\rm Z^0}$ to coupling dominantly to the electromagnetic current when $\mzp \ll M_{\rm Z^0}$~\cite{Kumar:2006gm, Baumgart:2009tn}.
 
The theory as it is has no predictive power over the mass spectrum of the fermions and scalars. But there is an interesting phase of the theory where the broad phenomenology becomes independent of the detailed mass spectrum. This happens when the fermions are heavier than the vector-boson, which is in turn much heavier than the scalars, $M_\Psi > \mzp > m_s$. In this case, we can integrate out the fermions and generate an effective coupling between the scalars and hypercharge~\cite{Steinberger-Schwinger}, 
\be
\label{eqn:sBB}
\mathcal{L}_{sBB} = \frac{1}{\Lambda_{\hp}} \hp B_{\mu\nu}B^{\mu\nu} + \frac{1}{\Lambda_{\Ap}} \Ap B_{\mu\nu}\tilde{B}^{\mu\nu}
\ee 
where $\hp$ stands for either the heavy or the light CP-even scalar, $\Ap$ is the pseudoscalar, and $\tilde{B}^{\mu\nu} = \epsilon^{\mu\nu\alpha\beta}B_{\alpha\beta}$ is the dual hypercharge field strength. The dimensionful coefficients are given by,
\be
\Lambda_{\hp,\Ap}^{-1} = \frac{\alpha_Y}{8\pi}\sum_f  \left(\frac{y_{\hp,\Ap}}{M_f}\right)  q_f^2~  A \left(\frac{m_{\hp,\Ap}^2}{4M_f^2}\right)
\ee  
where the sum runs over the heavy fermions. Here, $\alpha_Y = \alpha/\cos^2\theta_W$ is the hypercharge coupling, $y_{\hp,\Ap}$ is the Yukawa coupling of the respective scalar to the fermions, and the function $A(m_{\hp,\Ap}^2/4M_f^2) \rightarrow 4/3$ and $2$ in the limit $M_f \gg m_{\hp,\Ap}$ for the scalar and pseudoscalar, respectively (see ref.~\cite{Djouadi:2005gi} for the precise form of $A(x)$). Similar terms involving the $\zp$ as well as mixed hypercharge-$\zp$ are also generated with similar coefficients. 


The collider phenomenology of this phase is dominated by the production of the $\zp$ through the kinetic mixing, and its subsequent decay into the light scalars to which it couples directly. The scalars' subsequent decay into multiple photons therefore forms the distinctive signature of this type of models, and we now set to explore the detailed structure of such events. The total production cross-section is given approximately by
\be 
\label{eqn:LHCrate}
\sigma_{\rm tot} \approx 15~{\rm fb}~\left(\frac{\epsilon}{0.1}\right)^2 \left(\frac{1\TeV}{\mzp} \right)^5, 
\ee
where the scaling with the mass is approximate. There are strong generic bounds on kinetically mixed $\zp$ models, but for $\mzp \gtrsim$ a few hundred GeV the mixing parameter can be rather large $\epsilon\gtrsim 0.1$~\cite{Hook:2010tw}.   If only the pseudoscalar, $\Ap$, is lighter than the vector-boson then the $\zp$ may decay into $\Ap$+photon through the dimension-5 mixed operator $\Ap \tilde{\zp}_{\mu\nu}F^{\mu\nu}$. However, this decay mode is not parametrically much larger than the decay back to hypercharged SM fermions through kinetic mixing. These are independent operators, and whether one decay dominates over the other depends on the details of the high energy theory. We thus refrain from elaborating on this possibility further except to note two things. First, for the kinetic mixing contribution to the width of the $\zp$ to be smaller than the dimension-5 mixed operator contribution would require a kinetic mixing parameter of $\epsilon \lesssim 0.01$. As is clear from Eq.~(\ref{eqn:LHCrate}), this will result in too small a rate unless the $\zp$ is in the several hundreds GeV range and below. Second, the decay mode $\zp \rightarrow \Ap$+photon is best searched for in 3 photon searches if the pseudoscalar mass is not much smaller than the $\zp$ mass. Alternatively, when the pseudoscalar is very light, the diphoton resulting from its decay would form a photon-jet and result in the interesting new topology of photon + photon-jet. We henceforth concentrate on the other possibility where more than one scalar is light allowing for the decay $\zp\rightarrow \hp\Ap$, which would contain at least two pairs of diphotons.

Either way the scalars thus produced are unstable and will decay into diphotons. The  interactions~(\ref{eqn:sBB}) result in a decay of the scalars to two photons with relatively long lifetime. The boosted lifetime of the pseudoscalar for instance is given by
\be
\gamma c \tau_{\Ap} = 1~{\rm mm}\left(\frac{137^{-1}}{\alpha'}\right)\left(\frac{10\GeV}{m_{\Ap}}\right)^4\left(\frac{\mzp}{\TeV}\right)^3.
\ee
More importantly for the structure of such decays, the differential width for the decay of a scalar $s$, where $s$ stands for any of the scalars, with boost $\boost$, velocity $\beta = \sqrt{1-\boost^{-2}}$, and energy $E_s = \boost m_s$ in the lab frame is 
\be
\label{eqn:diffDecayWidth}
\frac{d^2\Gamma(s\rightarrow \text{diphotons})}{d\dimEn ~d\phi} = \frac{\Gamma_s}{2\pi \boost \beta}, 
\ee
where $\Gamma_s$ is the total decay width, $\phi$ is the angle of the production plane with respect to the scalar's momentum,  $\dimEn = 2E_2/m_s$ is the dimensionless energy of the less energetic photon, and $\boost(1- \beta) \le \dimEn \le \boost$. For the purpose of collider phenomenology, we will be especially interested in the behavior of the decay rate as a function of the separation $\Delta R = \sqrt{\Delta \phi^2 + \Delta \eta^2}$ between the two photons. This dependence can be worked out precisely, but the associated formulae are particularly simple in the large boost limit. For instance, the separation between the two photons is given by
\be
\label{eqn:DeltaR}
\Delta R = \frac{2\cosh\eta}{\sqrt{\dimEn \dimEn'}} + \mathcal{O}\left(\boost^{-1}\right), 
\ee
where $\eta$ is the rapidity of the scalar in the lab frame and $\dimEn' = 2\boost - \dimEn$ is the energy of the more energetic photon. Thus the separation of the two photons is greater than $\DeltaRmin$ when 
\be
0 < \dimEn <    \frac{2 \csc^2\theta}{\boost \Delta R_{\rm min}^2}. 
\ee
Since decays are evenly distributed in the photon's energy, $\epsilon$, the probability of the scalar decaying into two photons with $\Delta R > \DeltaRmin$ is simply given by 
\be
\label{labframe}
\frac{\Gamma\left(\Delta R > \DeltaRmin \right)}{\Gamma_s} = \frac{2  \cosh^2\eta}{\boost^2 \DeltaRmin^2}. 
\ee
A similar approximate formula can be obtained for the case when a minimum transverse energy cut is placed on the photons. The approximation above breaks down when 
\be
\boost \DeltaRmin  \cosh^{-1}\eta \lesssim 1
\ee 
but is otherwise very good. Importantly, in this boosted limit, Eq.~(\eqref{labframe}) is independent of the rapidity of the $\zp$ in the lab frame, but depends only on the rapidity of the scalars in the rest frame since $\boost^{-1} \cosh \eta = \hat{\boost}^{-1} \cosh\etahat$, where $\hat{\boost} = \mzp/2m_s$ and $\etahat$ are the scalar's boost and rapidity in the $\zp$ rest frame.

\begin{figure}[h]
\begin{center}
\includegraphics[width=\columnwidth]{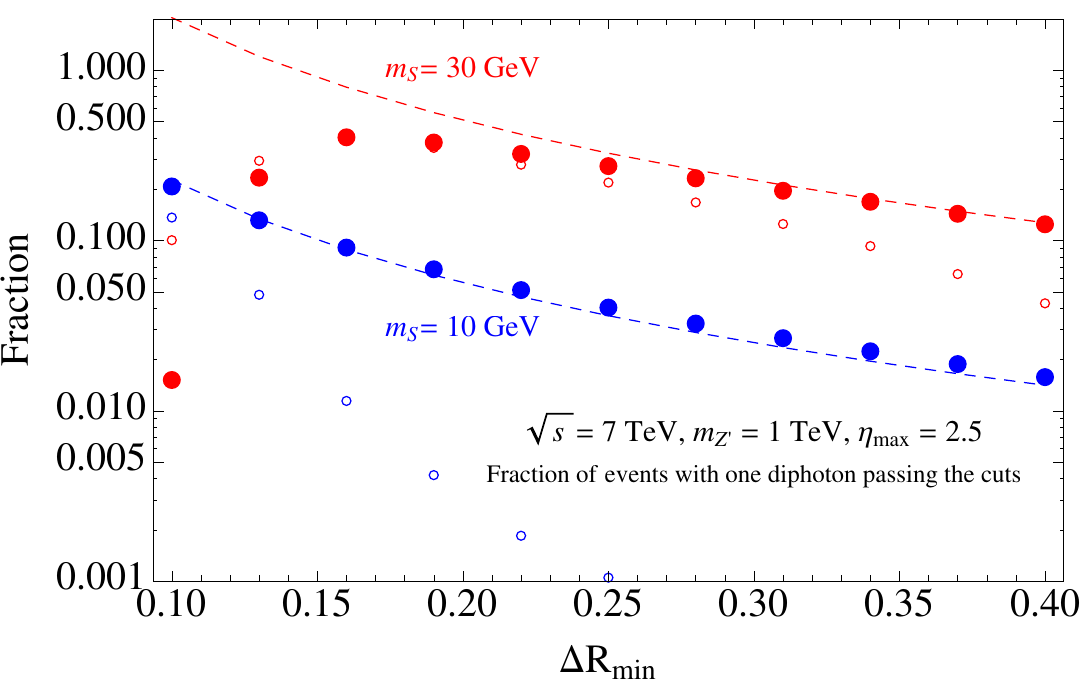}
\end{center}
\caption{The fraction of events with only one diphoton satisfying $\Delta R > \DeltaRmin$ for $\mzp = 1\TeV$ and two choices of the scalar mass. The dashed lines represent the approximate formula, Eq.~(\ref{eqn:frac1}) whereas the filled data points are from a monte-carlo simulation of the full event at the LHC with $\sqrt{s} = 7\TeV$. The hollow data points are simulation including a transverse energy cut on the photons of $10\GeV$. }
\label{fig:frac1}
\end{figure}

\begin{figure}[h]
\begin{center}
\includegraphics[width=\columnwidth]{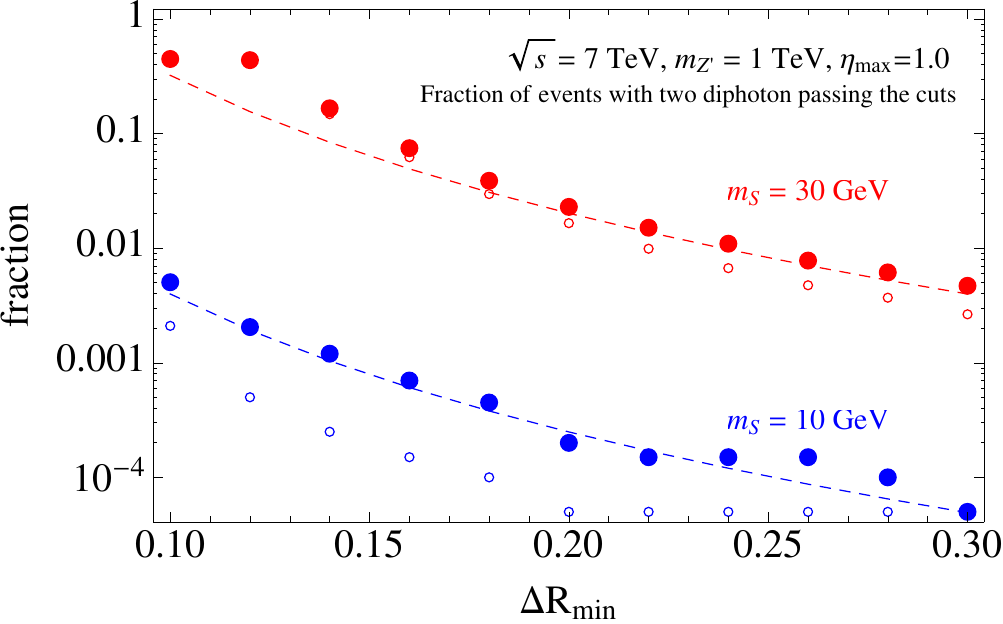}
\end{center}
\caption{The fraction of events with two diphotons both satisfying $\Delta R > \DeltaRmin$ for the same choice of parameters as in Fig.~\ref{fig:frac1}. We note the expected deterioration of the approximation, Eq.~(\ref{eqn:frac2}), as $\DeltaRmin$ diminishes as well as when the scalar mass increases resulting in a smaller boost. The loss of accuracy at higher $\DeltaRmin$ is only due to limited statistics.}
\label{fig:frac2}
\end{figure}

Computing more detailed observables requires accounting for geometric acceptance requirements and 
integrating over the full $\zp$ phase-space. Since the diphotons are fairly collimated, we can approximate the effect of a maximum detectable rapidity for photons, $\ymax$, by treating $\ymax$ as an upper limit on the scalars' rapidities. In the narrow-width approximation the leading order~\cite{Carena:2004xs} double differential cross-section for $\zp$ production depends on the $\zp$ lab frame rapidity, $y$, and the scalars rapidity in the $\zp$ rest frame, $\etahat$, as
\be
\frac{d^2\sigma}{dy~d\etahat} &=& \rho_{q,\bar{q}}(x_1,x_2) \times~\Theta(\etahat) ~\sigma_0 \\\nonumber &\times& {\rm BR}(Z'\rightarrow s_i \Ap),
\ee
where the parton luminosity function is $\rho_{q,\bar{q}}(x_1,x_2)$  with $x_{1,2} =\left(\mzp/\sqrt{s}\right) e^{\pm y}$,  $\sqrt{s}$ is the center of mass energy, 
\be
\sigma_0 =\frac{4\pi^2 ~\alpha \epsilon^2}{3M_{Z'}^2\cos^2\theta_{_W}}\left(\frac{Y_L^2+Y_R^2}{2} \right),
\ee
and $\Theta(\etahat)$ embodies the angular distribution of the $\zp$ decay products,
\be
\Theta(\etahat)=\frac{3}{4}\left(\frac{1-\tanh^2\etahat}{\cosh^2\etahat} \right). 
\ee
Here, the limits of integration on the lab frame rapidity are $|y| < \log\left(\sqrt{s}/\mzp\right)$. It is possible to integrate the full differential distribution including the decay of the scalar by applying the narrow-width approximation to the scalars. Specializing to the case $\zp \rightarrow \hp \Ap$, the fraction of events with four isolated photons can be computed exactly, yielding
\be
\label{eqn:frac2}
\frac{\sigma(\Delta R > \DeltaRmin)}{\sigma_{\rm tot}} &=& \frac{6\langle \ymax- y\rangle}
{{\hat \boost}_{\hp}^2{\hat\boost_{\Ap}^2}\Delta R_{\rm min}^4},
\ee 
with,
\be
\langle y\rangle = \frac{\sum_{q} \int dy~y~\sigma_0~ \rho_{q,\bar{q}}(x_1,x_2) }{\sum_{q} \int   dy~\sigma_0~ \rho_{q,\bar{q}}(x_1,x_2) }
\ee
where the limits of integration on the lab-frame rapidity were quoted above. Here we assumed $E^T_{\rm min} = 0$ for simplicity, but an analytic expression can be obtained also in the case of $E^T_{\rm min} \ne 0$. We note that the number of events with four isolated photons drops extremely rapidly with the rest-frame boost $\hat{\boost} = \mzp/2m_s$. Using the above expressions, we can also arrive at a simple approximate formula for the fraction of events where one diphoton pair is separated by less than $\DeltaRmin$, 
\be
\label{eqn:frac1}
\frac{\sigma(\Delta R_{1,2} \lessgtr \DeltaRmin)}{\sigma_{\rm tot}} &=& \frac{6\langle \tanh(\ymax-y)\rangle}{\hat{\boost}_*^{2}
\Delta R_{\rm min}^2 },%
\ee 
where $\hat{\boost}_*^{2} = 2/(\hat{\boost}_{\hp}^{-2}+\hat{\boost}_{\Ap}^{-2})$ is the harmonic mean of the two scalars' squared boosts.
In Figs.~\ref{fig:frac1} and \ref{fig:frac2} we plot the approximate analytic formulas \eqref{eqn:frac1} and \eqref{eqn:frac2} for the fraction of events with one (two) diphotons satisfying $\Delta R > \DeltaRmin$, together with results of monte-carlo simulation of the full event using the MadGraph4 package~\cite{Alwall:2007st}. In each case we have simplified to the case of equal scalar masses for simplicity, i.e. $m_{\hp} = m_{\Ap} = m_S$.

The above formulas make precise what is intuitively clear, namely that when the ratio of the $\zp$ mass to that of the scalar is large, $\mzp/m_s \gg 1$, most of the events will consist of two diphoton pairs, \emph{i.e.}~of two \textsl{photon-jets}. The above formulas are useful for quick and reasonable estimates for the fraction of phase-space where photon-jets are important in a variety of circumstances.  Depending on the detailed mass spectrum of the scalar sector, events may contain even more photon-jets. This occurs for example when the CP-odd scalar is the only light particle. In that case, events where the $\zp$ decays into $\hp\Ap$ (or $\Hp\Ap$) will result in three photon-jets as the CP-even scalars decay into two CP-odd ones ($\hp\rightarrow 2\Ap$ or $\Hp\rightarrow 2\Ap$). Such events are particularly interesting since they contain several distinct resonances. The angular distribution of the scalars in the event also carries important information, as is well known (e.g. \cite{AngularDistribution}). 


It remains to consider the experimental reconstruction of a photon-jet, in particular whether they would pass standard photon identification requirements, and how they may be discriminated from both individual photons and isolated neutral pions.  As a careful consideration requires full detector simulation, we restrict this discussion to several qualitative observations. The experimental signature of a photon-jet in the electromagnetic calorimeter (ECAL) may not be resolvable as either one or two standard isolated photon objects. Less boosted photon-jets produce photons that are physically separated at the ECAL by several Moliere radii, so that their showers will not overlap but the resulting photon objects will generically fail isolation requirements.  More boosted photon-jets result in overlapping showers from the two photons, producing a statistically broader $\eta-\phi$ profile in the electromagnetic calorimeter than normal photons. Even in this case, such objects will likely fail tight photon definitions, which are designed in part to reject the similar signal from decays of neutral pions and $\eta$'s. At ATLAS for example, the first layer of the electromagnetic calorimeter is very finely spaced, $\Delta\eta = 0.003$, which allows rejection of extremely boosted pions.  The above remarks apply to unconverted photons, but the conversion probability of photons in the tracker is significant. The double conversion of both photons may allow for a reconstruction of the photon pair invariant mass and serve as a discriminant against neutral pion decay. 

Thus, if the scalars are lighter than the $\zp$, the model discussed above would generically \emph{not} give conventional multi-photon signatures with well isolated photons.  One exception already noted is when the two photons in a photon-jet merge into a single reconstructed photon that passes quality cuts, giving rise to an apparent di-photon resonance of spin 1.  However, over most of the parameter space it is likely that a dedicated reconstruction algorithm would yield significantly higher efficiency.  High-mass diphoton events will also arise in events where one photon in each photon-jet is soft and/or wide-angle enough that its harder companion passes isolation requirements.  As can be seen from Figures 1 and 2, this happens with very low efficiency, but the subset of events with two isolated photons would appear as a broadly peaked signal in diphoton.   Present limits on Randall-Sundrum diphoton resonances at TeV masses, on the scale of $3-10$ fb \cite{Chatrchyan:2011fq}, only mildly constrain the parameter space of interest once the inefficiency of photon isolation for these signals is accounted for.

We close by mentioning several novel search possibilities. In the simplest case of intermediate mass scalars, where no strong collimation is expected, the particular model presented in this \textsl{Letter} motivates searches for a new $\zp$ resonance decaying into multiple photons. On the other hand, when the scalars, or more naturally only the pseudscalar, are much lighter than the $\zp$ mass, the majority of the diphoton pairs will be highly collimated and fail typical photon identification requirements. This phase of the theory prompts the consideration of a photon-jet as a distinct type of object in analogy with the recently proposed lepton-jets~\cite{ArkaniHamed:2008qp}. We stress that such $\zp$ need not be extremely heavy and searches for intermediate mass resonances are well-motivated. Finally, searches for doubly converted photons may be particularly powerful if the light resonance mass can be accurately reconstructed from the charged tracks and used as a discriminant against background.

%


\textbf{Acknowledgments:} We acknowledge helpful discussions with B. Brelier, K. Cranmer, Y. Gershtein, A. Haas, D. Krohn, J. Ruderman, P. Schuster, T. Spreitzer, and N. Weiner.

\end{document}